\documentclass[10pt,journal,compsoc]{IEEEtran}
\usepackage{amsmath}
\usepackage{amssymb}
\usepackage{booktabs}
\usepackage{multirow, booktabs,amssymb,amsmath}
\usepackage{soul,color,tabularx}
\usepackage{setspace}
\usepackage{xcolor,colortbl}
\usepackage{comment}
\usepackage[english]{babel}
\usepackage{graphicx}
\usepackage{epstopdf,subfig}
\usepackage{makecell}
\usepackage{algpseudocode}  
\usepackage{algorithmicx,algorithm}
\usepackage{cite}
\usepackage[square, comma, sort&compress, numbers]{natbib}
\usepackage{listings} 
\definecolor{dkgreen}{rgb}{0,0.6,0}
\definecolor{gray}{rgb}{0.5,0.5,0.5}
\definecolor{mauve}{rgb}{0.58,0,0.82}
\ifCLASSINFOpdf
\else
\fi
\begin{document}
%
\title{A Structure-aware Approach for Efficient Graph Processing}

\author{Beibei Si
\thanks{Manuscript received Month day, year; revised Month day, year.}}

\markboth{IEEE Transactions on Parallel and Distributed Systems,~Vol.~x, No.~x, Month~year}%
{Shell \MakeLowercase{\textit{et al.}}: Bare Demo of IEEEtran.cls for Computer Society Journals}

\IEEEtitleabstractindextext{%
\begin{abstract}
With the advent of the big data, graph are processed in an iterative manner, which incrementally described in the form of graph in big data applications. Most currently, graph processing methods treat the underlying map data as black boxes. However, as shown in experimental evaluation, graph structures often have diversity, different graph processing methods are very sensitive to the graph structure and show different performance for different data sets. Based on this, a graph processing method for graph structure analysis is proposed in this paper: (1) This paper calculates the vertex activity of a graph according to the in-degree and out-degree, and divide the corresponding vertices into the hot or cold partitions; (2) According to the change of graph structure caused by partial vertex convergence after iteration, this paper reclassifies the partitions, divides the lower active vertices into cold partition and reduces the frequency of calculation, which thereby reducing the cache miss rate and the I/O overhead caused by active vertices as well; (3) The partition with highest vertex status degree are given a priority calculation in this paper. In detail, more pronounced and more frequent vertices have higher processing priority. In this way, the convergence speed of the graph vertices is accelerated, and the running time of the graph algorithm in the big data environment is reduced. Our experiments show that compared with the latest system, the proposed method can double the performance of different graph algorithms and data sets.
\end{abstract}

}

\maketitle

\IEEEdisplaynontitleabstractindextext

%
\IEEEpeerreviewmaketitle
\IEEEraisesectionheading{\section{Introduction}\label{sec:introduction}}
\IEEEPARstart{I}{}ncrementally described in the form of graph in big data applications, graph are processed in an iterative manner. For example, search services (such as Google~\cite{Google}) use PageRank algorithm to sort results, social networks (such as Facebook~\cite{Facebook}) use Clustering algorithm to analyze user communities, knowledge sharing sites (such as Wikipedia~\cite{Wikipedia}) use Named Entity Recognition algorithm to identify text information, video sites (such as Netflix~\cite{Netflix} and Anysee~\cite{Anysee}) Based on Collaborative Filtering algorithm to provide film and television recommendations. Relevant studies indicate that the computational and storage characteristics of graph computing make it difficult for data-oriented parallel programming models to provide efficient support. The lack of description of correlation between data and inefficient support for iterative calculations can result in multiple times. Dozens of times the performance loss. The urgent need for an efficient Graph Computation system has made it one of the most important issues to be solved in the field of parallel and distributed processing. Current graph system processing strategy~\cite{GraphLab, PowerSwitch, HybirdGraph, Gemini, GraphChi, NXgraph, Mosaic} still lack of efficiency which listed below: (1) High cache miss rate; (2) Large I/O access overhead; (3) Slow convergence rate of large-scale graph data.

We profiled the solutions that resulted in the low performance of the existing representative graph systems.Due to the small-world phenomenon, the graph vertices will obey the power function distribution. A few graph vertices will connect the vast majority of graph vertices, while the vast majority of these vertices need to transfer state through these few vertices. Therefore, frequent visits and updates are needed for these core graph vertices while other vertices shortly converge, resulting in low frequency of access, thus confronting the problem mentioned above. So this paper adopts the graph partition of the dynamic increment, which will be explained explicitly in Section 3.

Currently, some work has already been done for graph partition of power law graph, but most of them are based on a distributed environment, regarding the underlying computing nodes as equivalent nodes. However, most graph processing methods treat the underlying graph data as black boxes, lacking research on dynamic graph partitioning and graph processing based on graph structure. However, in the real world, the graph structure is constantly changing. With iteration, a large number of graph vertices may converge in the graph partition. Frequent accesses to a small number of active vertices may result in repetitive loading of the entire graph partition including convergence, but these convergent vertices do not require access and processing, which leads to the severe waste of memory bandwidth and cache. The existing method does not consider the structural features of each partition, and the graph algorithm requires more update times for convergence and each update requires large overhead.

The graph vertex degree and its state degree have particularly critical influence on the convergence of graph vertices. Meanwhile, they also determine the processing order of the graph vertices. In the case of PowerSwitch system as shown in figure~\cite{PowerSwitch}, vertex 1 has a large degree and is more active. Theoretically, asynchronous method should be adopted to increase the convergence speed as a large number of graph vertices ($v_2,v_3,v_4,v_5$) require state transfer through active vertices. After updating its own data by asynchronous method, each vertex will be immediately updated through sending messages, so that the neighbors can be calculated by using the latest data. The vertices ($v_2,v_4,v_6$) have lower degree and will shortly converge, and it is of no high value to adopt asynchronous system to increase the convergence rate. The synchronization system should be adopted to reduce the cache miss rate and the time required for state updates of graph vertices.

Presently, the graph structure can be diverse, and its processing performance can be more different in a uniform way. Secondly, the graph structure formed by the unconverged graph vertices are constantly changing in operation, causing large fluctuations in performance. According to the above reasons, the paper proposes graph processing methods for graph structure perception. This paper incrementally obtain the graph structure characteristics formed by unconverged graph vertices in accordance with the analysis, adopting a suitable graph processing method for each graph partition block adaptively according to the underlying operation environment (the processor load, cache miss rate, etc. in each graph partition). More specifically, the main contributions of this work are summarized as follows:

\begin{itemize}
\item This paper analyzes the existing problems in the state-of-the-art distributed graph processing system and points out that the current graph processing system is lacked with targeted processing in the graph structure, affecting system performance.
\item This paper proposes the structure-centered graph partition and graph processing. According to the graph structure (graph vertices heat, etc.), the graph is partitioned by dynamic increment manner. The order of block partition is processed according to the graph schedule map of graph partition state degree.
\item This paper uses the graph structure perception combined with feature analysis in operation to switch each block of graph partition to the appropriate processing method.
\item The method is applied in the latest system. Experiments with five applications on five real-world graphs show that Gemini significantly outperforms existing distributed implementations, and the performance is improved by 2 times.
\end{itemize}

The remainder of this paper is organized as follows. Section $2$ analyzed the defects of the existing graph processing system, which puts forward the dynamic graph partitioning and adaptive graph processing optimization strategy. Section $3$ presents the dynamic graph partitioning modus, followed by adaptive graph processing method in Section $4$. Section $5$ shows experimental results. The related work is surveyed in Section $6$, and finally, Section $7$ concludes this work.
\section{Background and Motivation}
With the present of big data era, increasing data applications needed to be expressed in the form of vertices and edges, and processed through iterations. While state-of-the-art graph processing systems mainly concentrated on solving load balancing and communication overhead among varies of runtime environment, therefor ignoring the graph structural features of input data which have great impact on system performance. First, Assorted graph structure been processed may lead to immense performance differences with unified method; Second, Structure variations of the vertices that haven't converged in operation bring out volatile performance.

\begin{figure}[h]
\centering
\includegraphics[scale=0.25]{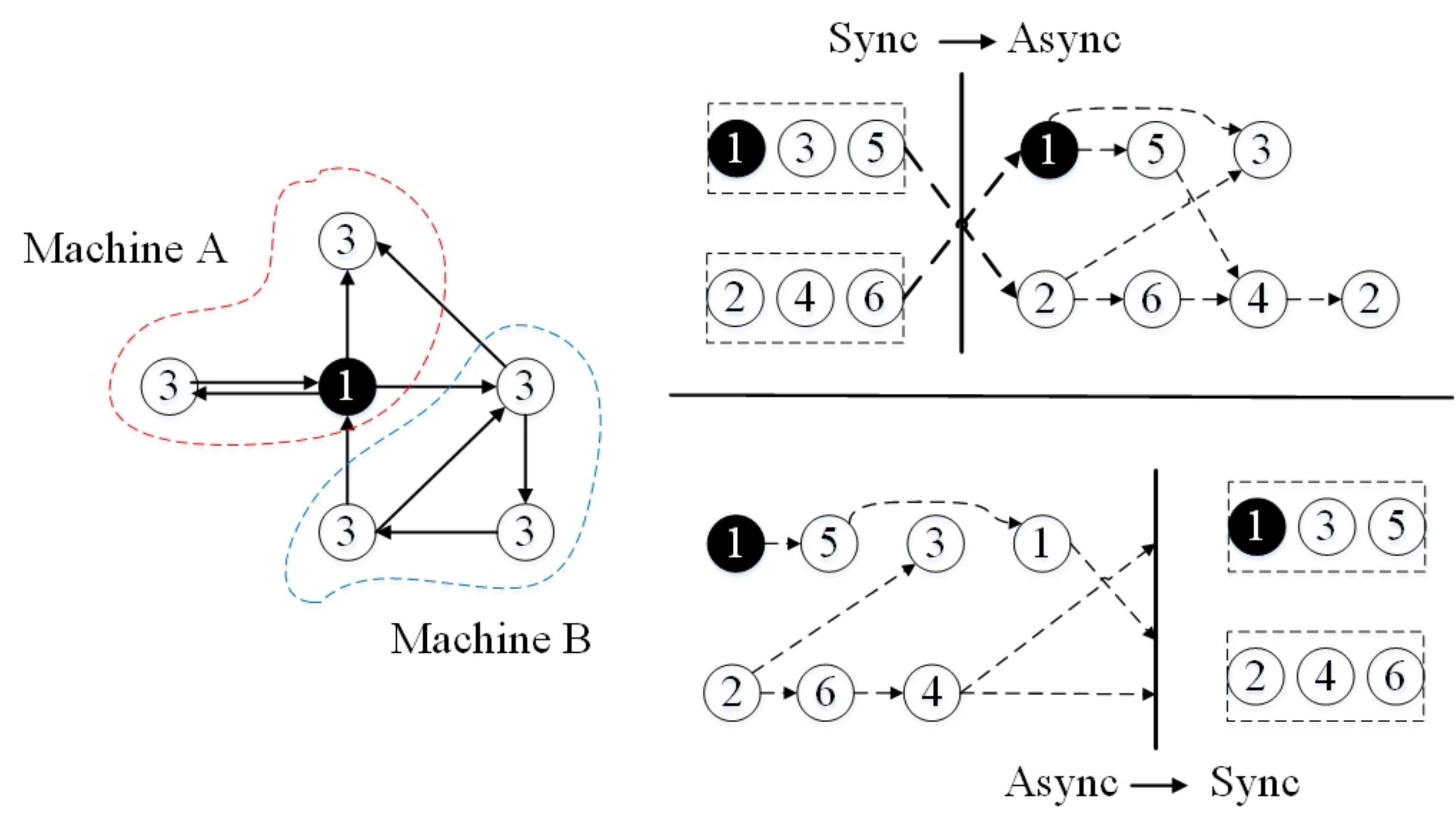}
\caption{Ineffective graph processing of partitions} 
\label{fig:PowerSwitch}
\end{figure}

Graph processing methods are very sensitive to the graph structure, therefore graph processing systems show quite different performance among diverse data sets. Most of the previous graph processing methods treats underlying data as black boxes, take neither partitioning nor processing strategy accordingly. Current graph computing model research work are mainly carried out in two aspects: one is focusing on performance optimization for a certain pattern, the other providing a same interface for two patterns(Synchronous mode and Asynchronous mode) that allows the user to choose according to the algorithmic features.Three issues has arisen due to the ignorance of above model, including low cache hit ratios, high input/output overhead, and slow convergence of large scale data.
\subsection{Disadvantages of Existing Methods}
To study the performance lose, We select some typical graph processing algorithms: PR (PageRank), CC (Connected Components), SSSP (Single-Source Shortest Paths), BFS (Breadth-First Search) and BC (Betweenness Centrality), along with commonly used graph data sets: amazon-2008, WikiTalk and twitter-2010 to evaluate the performance otherness among different algorithms and data sets. We set up experiments on an 8-node high-performance cluster interconnected with Infiniband EDR network (with up to 100Gbps bandwidth), each node containing two Intel Xeon E5-2670 v3 CPUs (12 cores and 30MB L3 cache per CPU) and 128 GB DRAM. We run 100 iterations on Gemini.

Figure~\ref{fig:6} shows the vertex convergence of six data sets with different structures under four algorithms through iterations. Figure~\ref{fig:6} gives detailed cache miss rate for different algorithms under different data sets. As shown in Figure~\ref{fig:6}, for the same data set, The structure of subgraph that non-convergent vertices composed of change continuously, traditional methods lack of the reflection to the diversity and dynamic changes of the graph structure, but a integrated graph partitioning and processing methods. The above strategies may depress convergence rate of the whole algorithm: In the iteration, some less active vertices have already converged while other remain active, which keep the entire partition loaded uninterruptedly and lead to decline in cache miss rate. (See Figure~\ref{fig:3})
\subsection{Optimized strategy}
We argue that inefficiency of traditional strategies mainly illustrated by following three points:

(1) Static graph partition methods. Structural diversification caused by vertex convergence is not considered. After one iteration. large number of vertices in each partition may converge, several vertices remain active, which result in frequent loading of a whole cache block, eventually wasting memory bandwidth, reducing cache hit rate, and frequent IO as well.

(2) Unified message processing mechanism. In terms of graph processing, The structural differences among graph partitions haven't been considered by the message passing model of existing systems, instead, they adopt a unified message processing mechanism. Some graph processing systems, such as $PowerSwitch$~\cite{PowerSwitch}, allow switching execution modes between synchronous and asynchronism, but are indistinguishably operated on all blocks. When synchronous message passing mechanism is adopted, the convergence speed of graph partition with more active vertices is limited. When asynchronous, High cache miss rate occur in partitions with less active vertices.

(3) Equal treatment to all graph partitions. The partitions are all treated the same as giving the same weight, nevertheless, It is known that natural graphs subject to skewed power-law degree distribution, which means small portion of vertex connects bulk of edge. Therefore, Frequent IO and high cache miss rate will arise in the event of average vertices partition.

For the reasons mentioned above, We present a novel graph structure-aware technique in the paper that obtains graph structure of the vertices that are not convergent by the analysis, and then incrementally partition the graph. After dynamic partition, We schedule the processing order of graph partitions, and for each iteration, adaptively choosing appropriate way to processing the graph partitions.In Summary, we have the following contributions:

\begin{itemize}
\item Our partition method separates the hot vertices from the cold, which endues the former with frequent update and significant change a higher priority, and reach the convergence faster, eventually reduce the average number of updates that an input graph needs to achieve converge.
\item After The graph partition s with dramatically drop-off in active vertices will be repartitioned after specific times of iterations. This method, on the one hand, takes the load balance problem caused by the change of graph structure into consideration, on the other hand, controls the computation overhead caused by the migration of vertices during dynamic graph partition.
\item We put high activity vertices with frequent updates into the same cache, for the vertices will be loaded in memory at the same time. By doing this, we reduce the overhead caused by inactive vertices and their loading times as well.
\end{itemize}
\section{Dynamic Graph Partition}
Due to the small-world phenomenon, the graph vertices will obey the power function distribution. A few graph vertices will connect with the vast majority of graph vertices, while the vast majority of these vertices need to transfer state through these few vertices. Therefore, frequent visits and updates are needed for these core graph vertices while other vertices rapidly reaching convergence, resulting in low frequency of access, thus confronting the problem mentioned above. Consequently, according to changes in graph structure caused by the convergence of some vertices during iteration. In this paper, partitions will be redivided, the less active vertices will be moved together to decrease the calculation frequency by graph partition manner of dynamic increment, thereby reducing the I/O overhead caused by active vertices and lowering the cache miss rate.
\begin{figure}[h]
\centering
\includegraphics[scale=0.18]{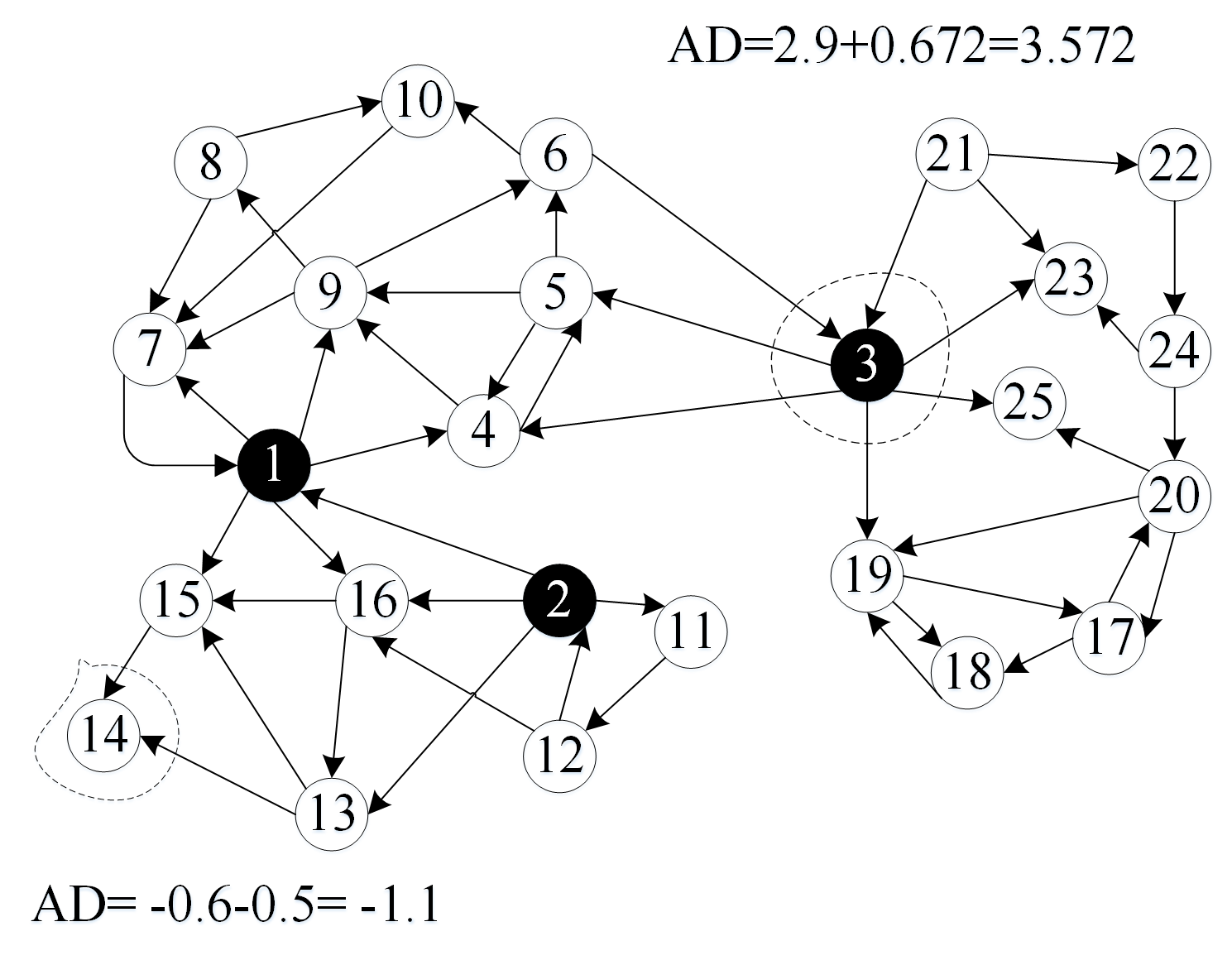}
\caption{Example graph} 
\label{fig:6}
\end{figure}
\subsection{Active Degree and State Degree}
Before getting to details, let us first give the targeted graph processing concepts. As the graph data increases dramatically, the researchers divided the graph data into several partitions and assigned the closely-related graph vertices to the same partition in order to accelerate convergence of the graph vertices. The input graph data is represented by $G = (V, E)$. While $V$ represents all the vertices and $E$ represents the edges of all the connected vertices. The current graph processing system stores the updated messages in the vertices by default, and the edges exist as fixed values. Therefore, the vertex degree is regarded as a fixed value in the computing.

\begin{table}
\begin{center}
\setlength\tabcolsep{1pt}
\begin{tabular}{cc}
\toprule  
Symbol&Definition\\
\midrule  
$D_i (v_i)$&In-degree of vertex $i$\\
$D_o (v_i)$&Out-degree of vertex $i$\\
$D(v_i)$&Degree function of vertex $i$\\
$D_{Max}(V)$&The maximum degree of all vertices\\
$SD(v_i)$&State degree of vertex $i$\\
$AD(v_i)$&Active degree of vertex $i$\\
$I_1$&Iteration that re-partitioning the partitions\\
$I_2$&Iteration that schedule cold partitions to compute\\
$T_1$&Threshold of vertices active degree\\
$T_2$&Threshold of vertices convergence\\
\bottomrule 
\end{tabular}
\caption{Definitions of symbols}
\label{fig:3}
\end{center}
\end{table}
{\bf Degree}\quad In this paper, $D_i (v_i)$ is used to represent in-degree of vertex $i$. The larger the in-degree, the more easily the vertex is affected by the neighbors. Which means, only when most neighbor vertex converge can the vertex tend to converge. Therefore, in the practical computation, vertices with large in-degree should be delayed to reduce the number of unnecessary updates. $D_o (v_i)$ indicates out-degree of vertex $i$. The greater the out-degree, more vertices will be affected by its update state. That indicates that only when the vertex converges can its neighbors tend to converge. Thence, in the practical computing, vertices with large out-degree should be processed in priority to accelerate the entire graph convergence. Regarding which mentioned above, the paper puts forward the concept of vertex power function, which is used to quantify the static structure features of graph vertices. Its formula is as follows:

\begin{align}
  D(v_i) = D_o(v_i) + \alpha*D_i(v_i)
\end{align}

The parameter $\alpha$ $($0.5$<\alpha<$1$)$ is an adjustable parameter, which is dynamically adjusted according to different data sets in the actual computation in order to achieve optimal performance. It can be a challenge to select the condition to match the value when computing heat value. The basis for selection is: value $α$ is adjusted according to the structure of input graph data. In the case of road network, a data set, each vertex has even in-edge and out-edge distributions and most graphs have similar vertex activity. The entire graph is of even distribution with value a trending to 0.5. However, in the case of a data set focused on by Weibo users, a few celebrities will have a large number of followers while most people have few followers, which leads to data skew. It amplifies the influence of vertex out-edge on the convergence of the entire graph, so value $\alpha$ will trend to 1 accordingly.

{\bf Active Degree}\quad The vertex activity depends not only on its degree function, but also on its neighbor vertex structure. In order to predict the initial activity information of each vertex in an input graph data set, the graph data is optimally partitioned under the condition of guaranteeing load balancing while improving the computing efficiency of subsequent iterations. The paper puts forward the structure features of quantification graph active degree, which are used as reference factors for the initial graph partition of data of data graph. It relies on the in-degree $D_i(v_i)$ and out-degree $D_o(v_i)$ of the vertex as well as the degree $D(v_k)$ of its neighbors.To this end, We use the hot-cold notion as in HotGraph and present our active degree algorithm function, scilicet the following $AD(v_i)$ :

\begin{equation}
  AD(v_i) = D(v_i) + \frac{\sum_{v_k}^{V} D(v_k)}{\sqrt{D_{Max}(V)} * D(v_i)}
\end{equation}

\begin{figure*}[!tb]
\centering
\includegraphics[scale=0.15]{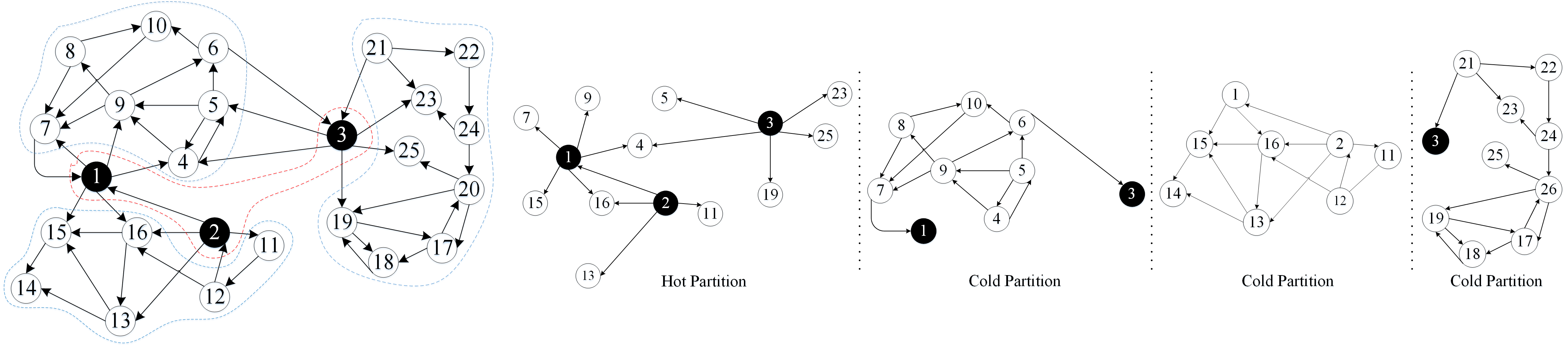}
\caption{Initial Chunk-based partitioning} 
\label{fig:6}
\end{figure*}

$D(v_k)$ indicates its neighbors’ degrees while $D_{Max}(V)$ indicates the maximum degree of all vertices, here we explore the feasibility of extending such design with fine-grained quantification of graph structure. The major difference is decoupling the in-degree and out-degree of vertices, note that unlike in HotGraph, $D(v_i)$ in this paper act like an degree function, taking in degree and out degree both into consideration and extending the graph data set to a more common directed graph.

$T_1$ is set as the active degree threshold, and is determined on the basis of user-defined sample size and the ratio of hot vertices, which follows T in HotGraph. For example: if the vertex number $V$ is viewed as 10000, the user-defined sample size is $V$=1000 and the ratio of hot vertex $R$ is 0.1, then the active degree threshold is $AD(V) = AD(v_{100})$, ie the active degree of the 100th vertex in the sample.

The vertices with active degree value $AD (v_i)$ greater than $AD(V)$ are marked as hot vertices and are stored in the hot partition. The vertices with active degree value smaller than $AD(V)$ are marked as cold vertices and stored in the cold partition. The hot and cold partitions are physically composed of cache blocks. The hot and cold vertices are stored in multiple cache blocks.  For instance, vertices with active degree value greater than 50 are hot vertices and the number is 200. On the contrary are cold vertices and the number is 2000. One cache block can store 100 vertices. Therefore, there are 2 hot partition and 20 cold partition. Particularly, vertices with 0 degree neither affect nor been affected by other vertices. Its convergence can be achieved in one iteration. This paper uniformly partitions them into regions with continuous addresses. The region is called as: dead partition.

{\bf State Degree}\quad According to the characteristics of input graph structure, $AD (V)$ evaluates the activity of vertices. As the vertices convergence, in iteration process, the activity would alter. Thereby, the state degree, $SD(v_i)$ represent the alteration of the activity of vertices in iteration process. The state degree means that when the state degree is higher, the state of graph vertices changed more. Also, more activity vertices have more influence on neighbor vertices. only when the vertex converges can its neighbors tend to converge. Otherwise, the low state degree vertices would continue to be updated. For different algorithmic, the definition of state degree and the methods of calculation are different, we will elaborate on the state degree formula corresponding to the common graph algorithm in section 3.3.

The partition state degree, $PSD(j)$ is the average of all vertices state degree accumulation in this partition. As a result of separation according to active degree value, the state degree of hot vertices is high and the state degree of cold vertices is low, which avoid this situation where low state degree vertices are more and there are fewer high state degree vertices so that the partition state degree improve. In conclusion, that the average of all vertices state degree accumulation in this partition is regarded as the state degree of the whole partition is reasonable.

The vertices state degree, $SD(v_i)$ and the partition state degree, $PSD(j)$ are applied in evaluating the activity of graph vertices and partition, respectively. In order to the whole graph can be convergence rapidly, as well as making high state degree vertices synchronous load to reduce cache invalidation, high state degree vertices and partition would be dealt priority.

Vertices active degree, $AD(v_i)$ and state degree, $SD(v_i)$ play an important role in vertices separation. This essay gives details about how to divide input graph structure and initial graph based in vertices activity in section 3.2 and section 3.3.
\subsection{Activity-based Partitioning}
In order to reduce the average number of updates that an input graph needs to achieve converge, this paper propose a graph partition strategy about graph structure sense, which not only is extensive, but also it has a characteristics that when the scale of data is larger, the performance is better. According to graph vertices in-degree and out-degree, vertices active degree, $AD(v_i)$ is calculated. And then according to $AD(v_i)$, graph vertices sort in descending. Based in this order, vertices are separated. The scale of partition is an exact of multiple of cache number. This separation act only is operated when data input. The time of reordering graph vertices is once in the whole algorithmic process. The expense produced by initial graph separation is divided to every time iteration. Not only it is helpful to improve cache hits rate, but also it lessen number of calculation. It can be proved that systems performance improvement is much more than extra expense. What’s more, for big scale input data, expense producing by every time become fewer contribution to great system extensiveness.

In initial iteration situation, 0 state degree vertices is investigated and put them into the dead partition. We Create a table named the first graph vertices degree table store vertices in-degree and out-degree. Moreover, We Create a table named the second graph vertices degree table to store position of neighbor vertices. The first graph vertices value table and the second graph vertices value table are applied in storing this time calculation value and last time calculation, respectively. Based in the value stored in the first graph vertices value table and the second graph vertices value table, vertices state degree and partition state degree can be known. To store partition ID and partition state degree, we create two tables, one is called ID table and the other one is partition state degree table. After this, we can separated hot partition and cold partition based in heat of vertices. As soon as all vertices are marked and separated to specific partitions, the table, partition state degree,is initialized and output initial partition.

\begin{algorithm}
\begin{footnotesize}
\caption{Initial Activity-based partitioning}
\label{algorithm:PNPFI}
\begin{algorithmic}[1]
\Procedure{Active\_Based Partition}{\emph{v$_i$, D$_o$(v$_i$),D$_i$(v$_i$)}}
\State expected chunk size $\leftarrow$ remain amount $/$ remain partitions
\While{ $V$ has unvisited vertex v$_i$}
\If{ \emph{D$_i$(v$_i$) = $0$ and D$_o$(v$_i$) = $0$}}
\State \emph{P$_{hot}$} $\leftarrow$ \emph{v$_i$}
\EndIf
\If{ \emph{AD(v$_i$) $\geqslant$ T$_1$}}
\State $hot\ edges$ $\leftarrow$ ${hot\ edges} \cup \emph{D$_o$(v$_i$)}$
\If{ ${hot\ edges}$ $>$ expected chunk size}
\State $hot\ partitions$ $\leftarrow$ $hot\ partitions$ $+\ 1$
\EndIf
\State \emph{hot\ partitions} $\leftarrow$ \emph{D$_o$(v$_i$)}
\State \emph{P$_{hot}$} $\leftarrow$ \emph{v$_i$}
\EndIf
\If{ \emph{AD(v$_i$) $\leqslant$ T$_1$}}
\State $cold\ edges$ $\leftarrow$ ${cold\ edges} \cup \emph{D$_o$(v$_i$)}$
\If{ ${cold\ edges}$ $>$ expected chunk size}
\State $cold\ partitions$ $\leftarrow$ $cold\ partitions$ $+\ 1$
\EndIf
\State \emph{cold\ partitions} $\leftarrow$ \emph{D$_o$(v$_i$)}
\State \emph{P$_{cold}$} $\leftarrow$ \emph{v$_i$}
\EndIf
\EndWhile
\EndProcedure
\end{algorithmic}
\end{footnotesize}
\end{algorithm}

Figure 5 gives an example of chunk-based partitioning, showing the vertex set on three nodes, with their corresponding dense mode edge sets. Knowing graph vertices active degree value and sorting them in descending relying on $AD(V)$, we separate graph vertices to two partition, $P_{cold}$ and $P_{hot}$. Each partition is made up of equal cache blocks. To read data conveniently, the scale of cache block is designed as the integral multiple of cache page number. For 0 state degree vertices, not only it does not received message of neighbor vertices, but also it can not transfer and update. And only one iteration can make it convergence. For this reason, we filter 0 state degree vertices firstly and deal with these data alone, which means these vertices would be separated when the state degree of vertices is calculated, would separate store them and would calculate priority in adaptive schedule period. When the iteration of 0 state degree vertices is achieved, there is no any act to reduce expense of iteration.

\begin{figure*}[!tb]
\centering
\includegraphics[scale=0.28]{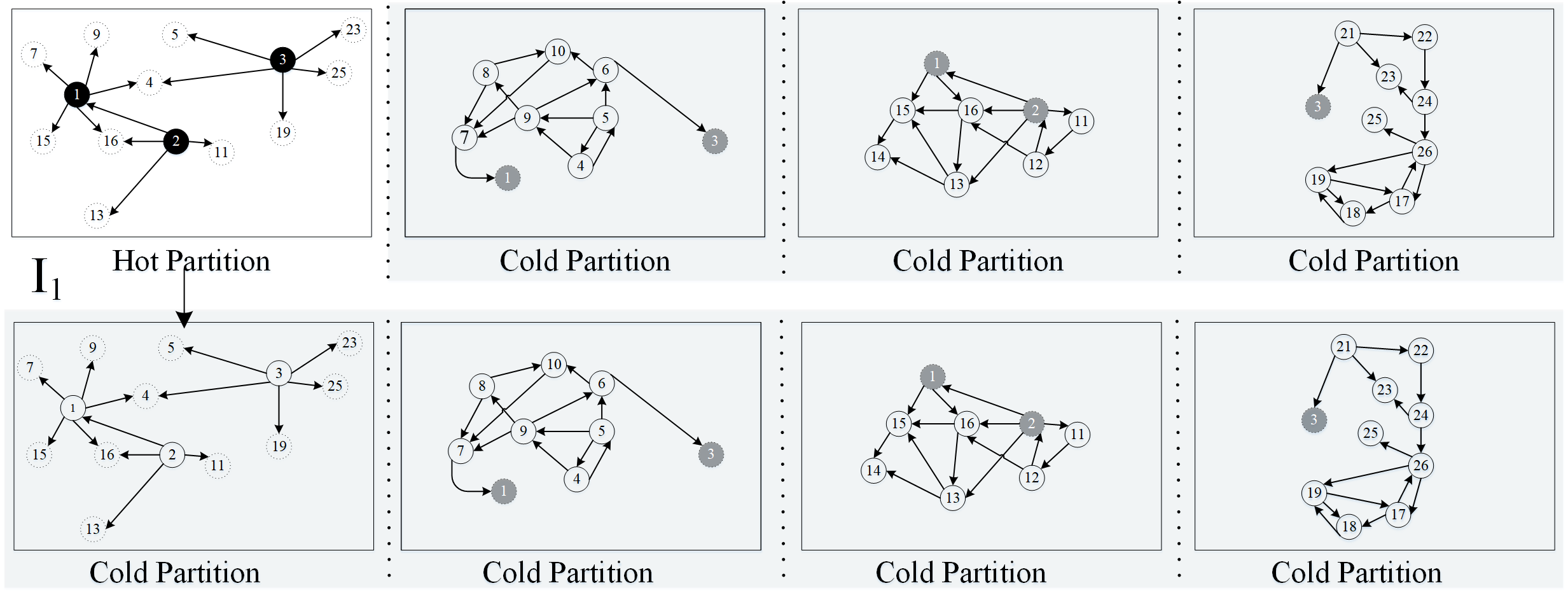}
\caption{Dynamic Structure-based graph partition for PageRank} 
\label{fig:7}
\end{figure*} 

\begin{figure}[h]
\centering
\includegraphics[scale=0.26]{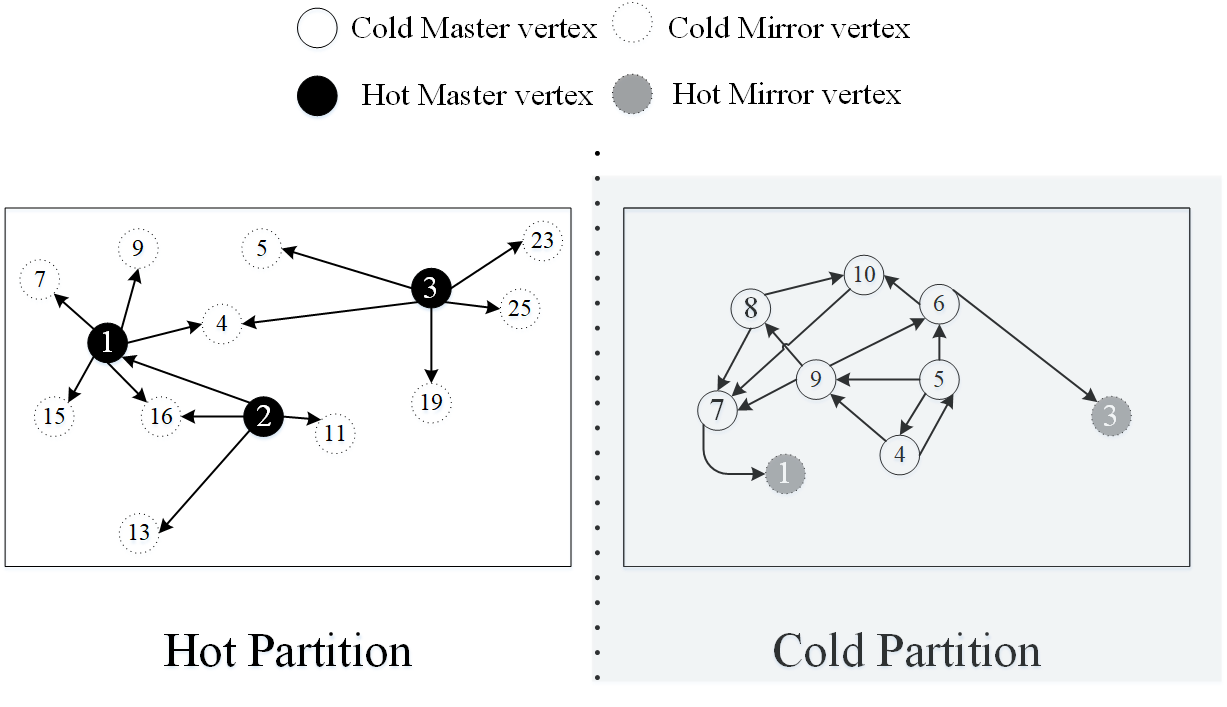}
\caption{The comparision of cold partition and hot partition} 
\label{fig:6}
\end{figure}

Due to the constant of edge data and input/output degree, we can preprocess input data and distinguish hot vertices and cold vertices relying on edge function, which is useful to increase rate of cache hits rate and decrease expense of I/O. Also, it is a great way to lessen number of iteration. Hot vertices become cold is a common trend in iteration process. There is a few cold vertices affected by neighbor vertices to be hot. It is essential to improve system performance which is separated based in heat graph partition. However, as the graph vertices convergence in calculation process, graph structure would modified so that it can not satisfy requirement that in initial partition, the high activity vertices is calculated priority during iteration. Because of this situation, dynamic increment graph partition is proposed. In initial partition, it would be separated again, according to graph vertices state.
\subsection{Structure-based Partitioning}
After a certain number of iteration, due to hot vertices convergence continuously, the number of hot vertices plumped. In order to make expense fall, hot partition would be rescheduled which would not result plenty of expense. Only to a marked variance is required to be updated. Time Complexity is $O(n)$.

The accumulation of the vertex state degree is obtained every $T_1$ iteration to obtain the average block state degree of all the hot and cold partitions and to determine whether there are hot partition with value smaller than the threshold $T_1$ and whether there are cold partition with value larger than threshold $T_1$. The hot partition with decreasing activity can be marked as the cold partition, and similarly, the cold partition with increasing activity can be marked as the hot partition. Because in the previous section, it is divided according to active degree value. Generally speaking, the state degree of hot vertices is higher while the state degree of the cold vertex is lower. So the phenomenon will not exist that many vertices with low state degree are in the partition while a few vertices with high state degree raise the state degree of the whole partition. Therefore, it is reasonable to use the average value of the vertex state degree in the partition as the state degree of the entire partition.

However, for some graph algorithms such as $PageRank$, the graph data shows the whole tendency from dense state to sparse state under these algorithms. The case fails to exist that the cold notion tan become the hot notion. In order to optimize the algorithm to reduce the program space occupation, the border variable barrier is maintained to partition cold and hot vertices. As the hot block gradually becomes cold, the barrier also moves accordingly. Compared with the universal partition method mentioned above which requires maintaining a tag variable table, the method only needs to maintain a $Vertex\_ID$ variable. However, for graph algorithms such as $SSSP$, the graph data tends to be dense and then tend to be sparse as a whole in these algorithms. That is to say, the cold vertices will first become hot and then converge, and a single barrier variable cannot represent the tendency. It requires the application of the universal method first proposed.

\begin{algorithm}
\begin{footnotesize}
\caption{Dynamic Structure-based Partition}
\label{algorithm:PNPFI}
\begin{algorithmic}[1]
\Function{Process\_Vertex}{\emph{v$_i$}, \emph{curr[ ]}, \emph{nexr[ ]}}
\State \emph{\#Pragma omp parallel reduction($+$:reducer)}
\While{ \emph{active vertices} all been visited}
\State \emph{local\_reducer} $\leftarrow$ \emph{local\_reducer} $+$ \emph{Process(v$_i$, curr[v$_i$], next[v$_i$])}
\State \emph{v$_i++$}
\EndWhile
\State \emph{reducer} $\leftarrow$ \emph{reducer} $+$ \emph{local\_reducer}
\State \emph{end Pragma}
\State \emph{global\_reducer} $\leftarrow$ \emph{global\_reducer} $+$ \emph{reducer}
\State \Return{\emph{global\_reducer}}
\EndFunction
\State
\Procedure{Structed\_Based Partition}{\emph{barrier}, \emph{curr$[$ $]$}, \emph{nexr$[$ $]$}}
\If{\emph{iteration} == \emph{I$_1$}}
\For{\emph{P$_{hot}$} and \emph{P$_{cold}$} have all been processed}
\For{\emph{v$_i$} belongs to \emph{Partition i}}
\State Process\_Vertex(\emph{v$_i$}, \emph{curr[ ]}, \emph{nexr[ ]})
\EndFor
\If{\emph{SD(P$_i$)} \emph{$<$} \emph{T$_1$} and \emph{P$_{hot}$}}
\State \emph{P$_{cold}$} $\leftarrow$ \emph{Partition i}
\State $barrier$ $\leftarrow$ $i$
\EndIf
\If{\emph{SD(P$_i$)} \emph{$>=$} \emph{T$_1$} and \emph{P$_{cold}$}}
\State \emph{P$_{hot}$} $\leftarrow$ \emph{Partition i}
\EndIf
\EndFor
\EndIf
\EndProcedure
\end{algorithmic}
\end{footnotesize}
\end{algorithm}

Figure~\cite{7} indicate the process of dynamic graph partitioning in $PageRank$. According to the accumulation state degree of each vertices, the average state degree of hot partition can be calculated. Find out the partition whose average state degree is less than $T_1$. And then, value of barrier is changed to be ID of the first vertices. This separation methods separate hot vertices again, but for cold vertices there is no effect. When it is calculated, hot vertices would fewer and fewer. Therefore, it is obvious that the scale of reschedule would zoom out.

When measuring the value of $PageRank$, the edge would be operated and divided, so the results relates to input edge and output edge of the vertex. The in-degree and the out-degree would straightly affected the vertices convergence. Hence, the difference of $Rank$ could be applied in evaluation the activity of vertices, which means, for $PageRank$, the definition of state degree is accumulation of the difference between this algorithmic result and last algorithmic result. Give $PageRank$ an example, assume that the first result is default 1 and there is no accumulation result at first time. If the second result is 5,and then it is obvious that the difference is 4. The accumulation is also 4. If the third result is 7, the difference between this result 7 and last result 5 is 2 so that the accumulation is 6, 4 and 2.

\begin{equation}
  \Delta_{PG} = \sum \left|Rank_{curr} - Rank_{next}\right|
\end{equation}

Figure~\cite{8} indicate the process of dynamic graph partitioning in $SSSP$. According to the accumulation state degree of each vertices, the average state degree of hot partition can be calculated. The partition whose average state degree is less than $T_1$, is marked as hot partition. Otherwise, it is marked as cold partition.

For $SSSP$, there is the same methods. When $SSSP$ was applied, because the calculation of the shortest path is relate to the accumulation account, it is not adaptable to evaluate the activity with the difference. In this methods, the smaller edge data between two calculation results is utilized, which is accumulated to decide whether there is modification of the vertices activity. Consequently, for $SSSP$, the definition of state degree is the accumulation of the smaller edge data between this result and last result. Same analogy to $CC$, it take a maximum clique. In this situation, the definition of state degree is the accumulation of the larger between this result and last result. The example of $CC$ is not a separate example here.

\begin{equation}
  \Delta_{SSSP} = min\{Edge\_data_{curr} , Edge\_data_{next}\}
\end{equation}


\begin{figure*}[!tb]
\centering
\includegraphics[scale=0.28]{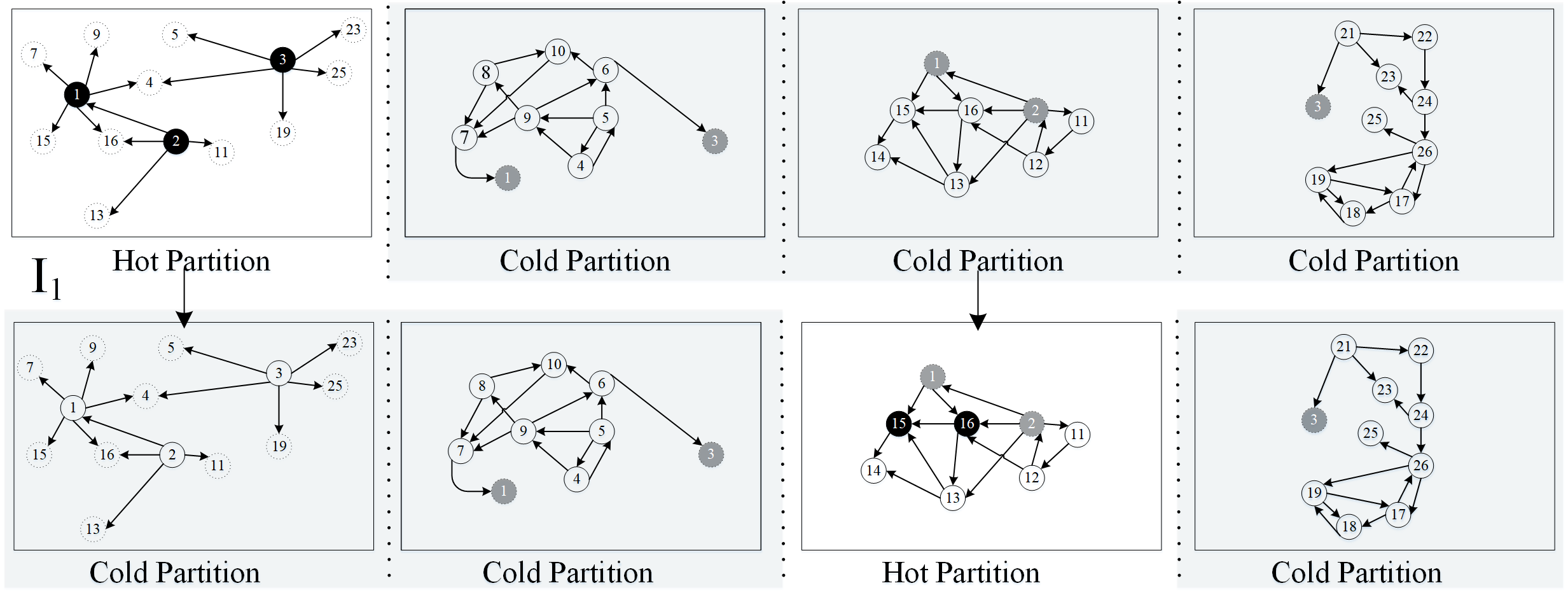}
\caption{Dynamic Structure-based graph partition for SSSP} 
\label{fig:8}
\end{figure*}

The number of dynamic Structure-based Partition is positively correlated with the number of iteration. For this reason, when the number of iteration goes up, the adjacent interval of rescheduling increases. Under the condition of ensuring the right results of algorithmic and avoiding extra expense, the rate of convergence is improved and the absence rate of cache is decreased.
\section{Adaptive Partition Scheduling}
In adaptive scheduling period, because of the different convergence rate of each vertices and the distinguish of the state degree of graph vertices, in calculating the accumulation of state degree of graph vertices process, the vertices that vary more frequent and greater have priority to be measured, as well as increasing the rate of convergence of graph vertices and shortening algorithmic running time. During $T_1$ iteration, hot partition would be separated again. After separation, if there is still hot partition, we adaptively scheduling hot partition and cold partition for calculation. If there is no situation where the whole graph is not convergence, the highest state degree cold partition is measured.

When there is hot partition after rescheduling, in each iteration process, the $n$ highest state degree cold partition and $m$ highest state degree hot partition are operated. The value of $m+n$ keep pace with the number of CPU. For example, if the number of CPU is 10, $m+n$ would be 10. For $I_2$ iteration, the value of $m$ and $n$ is decided by the algorithmic. Usually, it have to satisfy the condition $m>n$. It means that each time in hot partition the $m$ highest state degree cache partition are chosen and in cold partition the n highest state degree cache partition are chosen. On the contrary, if it is not $I_2$ iteration, we only apply the highest state degree hot partition. Thus, $n$ is equal to 0, and $m$ is equal to the number of CPU and equal to 10. Interval vertices stores with orders in ID sequence. If we need the specific partition to calculate, it represents reading in ascending ID order.

The sum of state degree values with all partitions is computed based on the partition state degree values stored in the partition state degree table. The smaller the state degree is, the closer the vertices are to convergence. When the sum of partition state degrees is smaller than a minimum value $T_2$, it can be regarded that the entire graph converges. Therefore, when the sum of state degree values with all partitions is smaller than the convergence threshold, it is determined that the entire picture converges, and the computation comes to an end and its result is output. Preferably, the specific value of the convergence threshold is defined by the user, and the default value is 0.000001.

According to a preferred mode of execution, the graph processing method stated further includes the step: judging whether it is the initial iteration. In the case of the first iteration, the block with the highest state degree in the hot partition is scheduled to be computed on the basis of computation the mentioned dead partition, and the convergence of the entire graph is determined after the iterative computations based on the sum of status degree value with all partitions. In the case that the entire graph does not converge, the subsequent iteration is proceeded.

\begin{algorithm}
\begin{footnotesize}
\caption{Adaptive Partition Scheduling}
\label{algorithm:PNPFI}
\begin{algorithmic}[1]
\Function{Procee\_Active}{\emph{m, n}, Partition \emph{P$_{hot}$, P$_{cold}$}}
\State \emph{threads} = \emph{numa\_num\_configured\_cpus()}
\For{each Partition \emph{p}}
\For{Vertex \emph{v$_i$} belongs to Partition \emph{p}}
\State SD(p) $\leftarrow$ SD(p) + Process\_Vertices(Process($v_i$), V)
\EndFor
\EndFor
\If{ Still remains \emph{P$_{hot}$}}
\If{iterations == I$_2$}
\State \emph{actives vertices} $\leftarrow$ \emph{m} * \emph{P$_{hot}$} + \emph{n} * \emph{P$_{cold}$}
\Else
\State \emph{actives vertices} $\leftarrow$ \emph{threads} * \emph{P$_{hot}$}
\EndIf
\EndIf
\If{ Only remains \emph{P$_{cold}$}}
\State \emph{actives vertices} $\leftarrow$ \emph{threads} * \emph{P$_{cold}$}
\EndIf
\State \Return{\emph{actives vertices}}
\EndFunction
\State
\Procedure{Scheduling}{\emph{active vertices}}
\For{Still remains untraversed Partition}
\State Send \emph{edge} in Partition \emph{p} to other nodes
\EndFor
\For{edge in Partition p hasn't all been received}
\State Receive \emph{edge} in Partition \emph{p} from other nodes
\EndFor
\If{ \emph{P$_{hot}$}}
\State master $\leftarrow$ mirror vertex update
\EndIf
\If{ \emph{P$_{cold}$}}
\State mirror $\leftarrow$ master vertex update
\EndIf
\EndProcedure
\end{algorithmic}
\end{footnotesize}
\end{algorithm}

\begin{figure}[tb]
\centering
\includegraphics[scale=0.36]{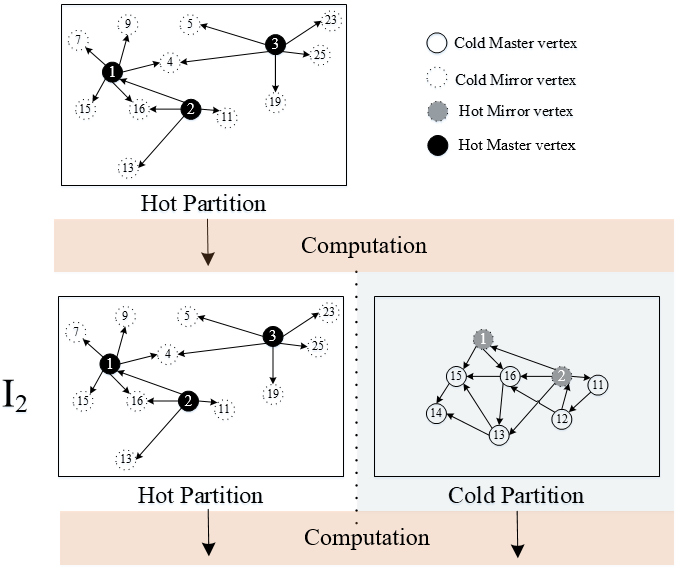}
\caption{Adaptive Partition Scheduling} 
\label{fig:9}
\end{figure}
One of the challenges of adaptive scheduling is to ensure that the hot partition is sufficiently computed. When the number of hot partitions is greater than that of machine threads, it indicates that one single iteration fails to make all hot partitions to be computed. Therefore, it is necessary to ensure for scheduling of partition that the hot partition with higher heat can be computed after the activity of the hot partition is reduced. It should be noted that it is a long process when the hot partition is repeatedly computed and the activity declines, trending to the activity of the cold partition. Due to the complexity of the structure, even the computation number increases, the convergence condition the hot partition requires is also more than that of the cold partition. And when all the hot partitions tend to converge, the entire graph will tend to converge and the graph algorithm will also be close to the end of computation.

In addition, regarding the convergence threshold: (1) Whether the algorithm actually converges or whether the computation is completed has nothing to do with the convergence threshold. $T_2$ is just a value for judging whether the current algorithm converges as the accomplish time of computation cannot be known in operational process of algorithm. Therefore, the state degree should be obtained at regular intervals to obtain the result that if it has converged compared with $T_2$. Consequently, it is not the case that the smaller the D2 is set, the faster the algorithm converges. (2) It does not make much sense that different convergence thresholds should be set based on different algorithms or application cases. Because the state degree of all algorithms can be 0 when reaching the convergence. It requires a relatively long time from 0.000001 to 0.  In order to improve performance, the state degree 0.000001 is considered as convergence, and the convergence threshold is fixed at 0.000001 while the algorithm result is within the tolerance range.

\section{Related Work}
Previous graph processing systems, over distributed~\cite{Pregel, X-Pregel, GraphLab, PowerGraph, Giraph, PowerSwitch, HybirdGraph, Gemini} or single multi-core platform~\cite{GraphChi, TurboGraph, VENUS, GridGraph, NXgraph, Mosaic}, have done plenty of work on effective graph processing such as load balancing and communication overhead reducing. Most of those approaches treat graph data as black box, which means, graph have been managed as a combination of vertices and edges (i.e.Vertex-centric and Edge-centric) rather than logical structure, the difference among which generates performance variability. In this section, we give a brief summary of related categories of prior work.

\textbf{Distribute Systems:} Pregel~\cite{Pregel} divide the graph by hashing the vertex id which ensure loading balancing. Yet Pregel uses message communication paradigm, messages that need to be processed will be huge when vertices has many adjacent points. Coincidentally, Pregel performs inefficiency in power-law graph and only allows global synchronization. X-Pregel~\cite{X-Pregel} optimizes Pregel's messaging mechanism by reducing the number of messages that needed to be delivered in every iteration. Giraph~\cite{Giraph} adds more features compare to Pregel, including master computation, out-of-core computation, etc. But the poor locality of data access limits its effective.

GraphLab~\cite{Giraph} follows the vertex-centric GAS model, but its partitions are still obtained by randomly division. On the other hand, Its shared memory storage strategy may have performance bottlenecks for large graphs. PowerGraph~\cite{PowerGraph} works well on power-law graph, but no special optimizations are considered for speeding up I/O access just as GraphLab. PowerSwitch~\cite{PowerSwitch} proposed an adaptive graph processing method based on PowerGraph, adaptively switching between synchronous and asynchronous processing modes according to the amount of vertices processed per unit time to achieve the best performance. However, it treats all the vertices of a graph as the same, and does not handle the convergence according to the vertices in the iteration. PREDIcT~\cite{PREDIcT} proposes an experimental methodology for predicting the runtime of iterative algorithms, which optimizes cluster resource allocations among multiple workloads of iterative algorithms.

Maiter~\cite{Maiter} propose delta-based accumulative iterative computation which reduce costs and accelerate calculations.HybridGraph~\cite{HybirdGraph} puts forward a algorithm adaptively switching between pull and push, focusing on performing graph analysis on a cluster IO-efficiently. Compare to GraphLab, PowerGraph employs a vertex-cut mechanism to reduce the network cost of sending requests and transferring messages at the expense of incurring the space cost of vertex replications. GrapH~\cite{GrapH} focus on minimize overall communication costs by using an adaptive edge migration strategy to avoid frequent communication over expensive network links. Gemini~\cite{Gemini} is a computation-centric distributed graph processing system that uses a hybrid pull/push approach to facilitate state updates and messaging of graph vertices.

\textbf{Single-machine Systems:} GraphChi~\cite{GraphChi} is a vertex-centric graph processing system and improve IO access efficiency by parallel Sliding Window processing strategy. But the outgoing edges of all vertices have to be loaded into memory before computation, resulting in unnecessary transfer of disk data. Also, all memory blocks have to be scanned when accessing neighboring vertices, which lead to inefficient graph traversal. TurboGraph~\cite{TurboGraph} proposed a Pin-And-Slide model to solve this problem. PAS has no delay in dealing with local graph data, but only applies to some specific parallel algorithms. Compare to the two above, VENUS~\cite{VENUS} expands to nearly every algorithm and enables streamlined processing which performs computation while the data is streaming in. Moreover, it uses a fixed buffer to cache the v-shard, which can reduce random IO.

GridGraph~\cite{GridGraph} uses a 2-level Hierarchical Partitioning scheme to reduce the amount of data transfer, enable streamlined disk access, and maintain locality. But it requires more disk data transfer using TurboGraph-like updating strategy. Besides, it cannot fully utilize the parallelism of multi-thread CPU without sorted edges. NXgraph~\cite{NXgraph} propose the Destination-Sorted Sub-Shard (DSSS) structure to store graph with three updating strategies: SPU, DPU and MPU. it adaptively choose suitable one to fully utilize the memory space and reduce the amount of data transfer. It achieves higher locality than v-shards in VENUS~\cite{VENUS} and reduces the amount of data transfer and enables streamlined disk access pattern. Mosaic~\cite{Mosaic} combines fast host processors for concentrated memory-intensive operations, with coprocessors for compute and I/O intensive components.

Traditional graph systems, either memory-share nor distribute, take the variable of graph structure into consideration, which appears through constantly convergence of vertices during iterations and plays significant role in program optimization. In this case, We present a novel graph structure-aware technique in the paper that provide adaptive graph partitioning and processing scheduling according to the variety of graph structure. Our strategy reduce the overhead caused by inactive vertices and their loading times as well speed up convergence rate.
\section{Conclusion}
In this paper, We adopted a structure-centric distributed graph processing method, Through graph structure perception, graph structure features of unconvergent vertices are incrementally obtained according to analysis, adaptively scheduling suitable graph processing methods. Our development reveal that (1) The dynamic incremental partitioning of vertex degree and state degree can significantly reducing IO resource overhead and cache miss rate, and (2) Computation and communication overhead of less active vertices can be reduced by setting priority of graph partitions and scheduling them based on predestinated order, and accelerated the algorithm convergence as well. Our experimental results on a variety of different data sets and their structural features of the graph demonstrate the efficiency, effectiveness and scalability of our approach, in comparison to state-of-the-art race detection approaches.

\bibliographystyle{IEEEtran}
\bibliography{references}

\begin{thebibliography}{10}
\providecommand{\url}[1]{#1}
\csname url@samestyle\endcsname
\providecommand{\newblock}{\relax}
\providecommand{\bibinfo}[2]{#2}
\providecommand{\BIBentrySTDinterwordspacing}{\spaceskip=0pt\relax}
\providecommand{\BIBentryALTinterwordstretchfactor}{4}
\providecommand{\BIBentryALTinterwordspacing}{\spaceskip=\fontdimen2\font plus
\BIBentryALTinterwordstretchfactor\fontdimen3\font minus
  \fontdimen4\font\relax}
\providecommand{\BIBforeignlanguage}[2]{{%
\expandafter\ifx\csname l@#1\endcsname\relax
\typeout{** WARNING: IEEEtran.bst: No hyphenation pattern has been}%
\typeout{** loaded for the language `#1'. Using the pattern for}%
\typeout{** the default language instead.}%
\else
\language=\csname l@#1\endcsname
\fi
#2}}
\providecommand{\BIBdecl}{\relax}
\BIBdecl

\bibitem{Google}
``Google,'' \url{http://www.google.com/}, 2018.

\bibitem{Facebook}
``facebook,'' \url{http://www.facebook.com/}, 2018.

\bibitem{Wikipedia}
``Wikipedia,'' \url{http://zh.wikipedia.org/}, 2018.

\bibitem{Netflix}
``Netflix,'' \url{http://www.netflix.com/}, 2018.

\bibitem{Anysee}
X.~Liao, H.~Jin, Y.~Liu, L.~M. Ni, and D.~Deng, ``Anysee: Peer-to-peer live
  streaming,'' in \emph{Proceedings of the IEEE International Conference on
  Computer Communications}.\hskip 1em plus 0.5em minus 0.4em\relax IEEE, 2006,
  pp. 1--10.

\bibitem{GraphLab}
Y.~Low, D.~Bickson, J.~Gonzalez, C.~Guestrin, A.~Kyrola, and J.~M. Hellerstein,
  ``Distributed graphlab: a framework for machine learning and data mining in
  the cloud,'' \emph{Proceedings of the VLDB Endowment}, vol.~5, no.~8, pp.
  716--727, 2012.

\bibitem{PowerSwitch}
C.~Xie, R.~Chen, H.~Guan, B.~Zang, and H.~Chen, ``Sync or async: Time to fuse
  for distributed graph-parallel computation,'' in \emph{ACM SIGPLAN Notices},
  vol.~50, no.~8.\hskip 1em plus 0.5em minus 0.4em\relax ACM, 2015, pp.
  194--204.

\bibitem{HybirdGraph}
Z.~Wang, Y.~Gu, Y.~Bao, G.~Yu, and J.~X. Yu, ``Hybrid pulling/pushing for
  i/o-efficient distributed and iterative graph computing,'' in
  \emph{Proceedings of the 2016 International Conference on Management of
  Data}.\hskip 1em plus 0.5em minus 0.4em\relax ACM, 2016, pp. 479--494.

\bibitem{Gemini}
X.~Zhu, W.~Chen, W.~Zheng, and X.~Ma, ``Gemini: A computation-centric
  distributed graph processing system.'' in \emph{OSDI}, 2016, pp. 301--316.

\bibitem{GraphChi}
A.~Kyrola, G.~E. Blelloch, and C.~Guestrin, ``Graphchi: Large-scale graph
  computation on just a pc,'' in \emph{OSDI}.\hskip 1em plus 0.5em minus
  0.4em\relax USENIX, 2012.

\bibitem{NXgraph}
Y.~Chi, G.~Dai, Y.~Wang, G.~Sun, G.~Li, and H.~Yang, ``Nxgraph: An efficient
  graph processing system on a single machine,'' in \emph{Data Engineering
  (ICDE), 2016 IEEE 32nd International Conference on}.\hskip 1em plus 0.5em
  minus 0.4em\relax IEEE, 2016, pp. 409--420.

\bibitem{Mosaic}
S.~Maass, C.~Min, S.~Kashyap, W.~Kang, M.~Kumar, and T.~Kim, ``Mosaic:
  Processing a trillion-edge graph on a single machine,'' in \emph{Proceedings
  of the Twelfth European Conference on Computer Systems}.\hskip 1em plus 0.5em
  minus 0.4em\relax ACM, 2017, pp. 527--543.

\bibitem{Pregel}
G.~Malewicz, M.~H. Austern, A.~J. Bik, J.~C. Dehnert, I.~Horn, N.~Leiser, and
  G.~Czajkowski, ``Pregel: a system for large-scale graph processing,'' in
  \emph{Proceedings of the 2010 ACM SIGMOD International Conference on
  Management of data}.\hskip 1em plus 0.5em minus 0.4em\relax ACM, 2010, pp.
  135--146.

\bibitem{X-Pregel}
N.~T. Bao and T.~Suzumura, ``Towards highly scalable pregel-based graph
  processing platform with x10,'' in \emph{Proceedings of the 22nd
  International Conference on World Wide Web}.\hskip 1em plus 0.5em minus
  0.4em\relax ACM, 2013, pp. 501--508.

\bibitem{PowerGraph}
J.~E. Gonzalez, Y.~Low, H.~Gu, D.~Bickson, and C.~Guestrin, ``Powergraph:
  distributed graph-parallel computation on natural graphs.'' in \emph{OSDI},
  vol.~12, no.~1, 2012, p.~2.

\bibitem{Giraph}
C.~Avery, ``Giraph: Large-scale graph processing infrastructure on hadoop,'' in
  \emph{Proceedings of the Hadoop Summit. Santa Clara}, vol.~11, no.~3, 2011,
  pp. 5--9.

\bibitem{TurboGraph}
W.-S. Han, S.~Lee, K.~Park, J.-H. Lee, M.-S. Kim, J.~Kim, and H.~Yu,
  ``Turbograph: a fast parallel graph engine handling billion-scale graphs in a
  single pc,'' in \emph{Proceedings of the 19th ACM SIGKDD international
  conference on Knowledge discovery and data mining}.\hskip 1em plus 0.5em
  minus 0.4em\relax ACM, 2013, pp. 77--85.

\bibitem{VENUS}
J.~Cheng, Q.~Liu, Z.~Li, W.~Fan, J.~C. Lui, and C.~He, ``Venus: Vertex-centric
  streamlined graph computation on a single pc,'' in \emph{Data Engineering
  (ICDE), 2015 IEEE 31st International Conference on}.\hskip 1em plus 0.5em
  minus 0.4em\relax IEEE, 2015, pp. 1131--1142.

\bibitem{GridGraph}
X.~Zhu, W.~Han, and W.~Chen, ``Gridgraph: Large-scale graph processing on a
  single machine using 2-level hierarchical partitioning.'' in \emph{USENIX
  Annual Technical Conference}, 2015, pp. 375--386.

\bibitem{PREDIcT}
A.~B. Adrian Daniel~Popescu, ``Towards predicting the runtime of iterative
  analytics with predict,'' in \emph{Ecole Polytechnique Federal de
  Lausanne}.\hskip 1em plus 0.5em minus 0.4em\relax EPFL, 2013.

\bibitem{Maiter}
L.~G. Yanfeng~Zhang, Qixin~Gao, ``Maiter: An asynchronous graph processing
  framework for delta-based accumulative iterative computation,'' in \emph{IEEE
  transactions on parallel and distribution and distrbuted systems}.\hskip 1em
  plus 0.5em minus 0.4em\relax IEEE, 2014, pp. 2091--2100.

\bibitem{GrapH}
M.~A.~T. Christian~Mayer, ``Graph: Heterogeneity-aware graph computation with
  adaptive partitioning,'' in \emph{2016 IEEE 36th International Conference on
  Distributed Computing Systems}.\hskip 1em plus 0.5em minus 0.4em\relax IEEE,
  2016.

\end{thebibliography}
\end{document}